\documentclass[pra,preprint,tightenlines,showpacs,11pt]{revtex4}\usepackage{epsfig} 
     
\begin{document} 

\title{ 
Imaging Molecules from Within: Ultra-fast, {\AA}ngstr\"om 
Scale Structure Determination of Molecules 
via Photoelectron Holography using Free Electron Lasers } 

\author{ F. Krasniqi$^1$, B. Najjari$^2$, L. Str\"uder$^{3,1}$, D. Rolles$^1$,  
A. Voitkiv$^2$ and J. Ullrich$^{1,2}$} 
\affiliation{ $^1$ Max Planck Advanced Study Group, Center for 
Free Electron Laser Science, Hamburg, Germany \\ 
$^2$ Max-Planck-Institut f\"ur Kernphysik, 
Saupfercheckweg 1, D-69117 Heidelberg, Germany \\ 
$^3$ MPI Halbleiterlabor, Otto-Hahn-Ring 6, D-81739 M\"unchen, Germany } 
\begin{abstract} 

A new scheme based on (i) upcoming brilliant X-ray Free Electron Laser (FEL) sources, 
(ii) novel energy and angular dispersive, large-area electron imagers and 
(iii) the well-known photoelectron holography is elaborated that provides 
time-dependent three-dimensional structure determination of small to medium 
sized molecules with {\AA}ngstr\"om spatial and femtosecond time resolution. 
Inducing molecular dynamics, wave-packet motion, dissociation, passage through 
conical intersections or isomerization by a pump pulse this motion is visualized 
by the X-ray FEL probe pulse launching keV photoelectrons within few femtoseconds 
from specific and well-defined sites, deep core levels of individual atoms, 
inside the molecule. On their way out the photoelectrons are diffracted generating 
a hologram on the detector that encodes the molecular structure at the instant 
of photoionization, thus providing 'femtosecond snapshot images of the molecule from within'. 
Detailed calculations in various approximations of increasing sophistication 
are presented and three-dimensional retrieval of the spatial structure 
of the molecule with {\AA}ngstr\"om spatial resolution is demonstrated. 
Due to the large photo-absorption cross sections the method extends 
X-ray diffraction based, time-dependent structure investigations envisioned 
at FELs to new classes of samples that are not accessible by any other method. 
Among them are dilute samples in the gas phase such as aligned, oriented or 
conformer selected molecules, ultra-cold ensembles and/or molecular or 
cluster objects containing mainly light atoms that 
do not scatter X-rays efficiently. 

\end{abstract} 


\maketitle 


\section*{Introduction}

The vision to directly follow time-dependent structural changes 
when molecular bonds are formed or break apart, when transition states 
or conical intersections are passed, namely to 'make the molecular movie' 
on atomic i.e. sub-nanometer length and molecular i.e. femtosecond time scales 
is among the strongest motivations driving huge efforts worldwide 
to develop next generation light sources, 
the Free Electron Lasers (FEL) \cite{1}-\cite{5}. 
Delivering ultra-intense bursts of $10^{13}$ coherent photons at up to 
$12$ keV energies with pulse lengths of $10$ to $100$ fs and realistic 
prospects to reach $1$ fs in the future, the 'standard' scenario 
is to extract the (time-dependent) structure of molecules in the gas 
phase via coherent X-ray diffraction. Focusing the X-rays to spots 
as small as $100$ nm onto single objects this extends to the hope 
to image individual molecules in the gas phase \cite{6}. 
This would, in principle, enable (time-dependent) 
structure determinations of many non-crystalizable molecules 
of biological interest, considered to represent a major break-through 
in structural biology.

Nevertheless, the realization of these visions is by far 
not assured essentially due to two reasons. First, even though 
the photon flux is huge, the tiny X-ray diffraction cross section 
of typically $\sim 10^{-24}$ cm$^2$ results in just a few tens 
to a few thousands of scattered photons in one shot from e.g. 
an aligned gas phase molecular ensemble or from a large, 
single molecule, respectively. Thus, thousands of individual shots 
with known relative orientation of the molecules have to be summed 
in order to obtain the statistical significance requested for atomic 
spatial resolution. Up to now there is no pathway identified without 
controversy to reach that goal. Second, the 'destructive' photo 
absorption cross section is factors of ten (e.g. for the carbon K-shell 
in bio-molecules) to thousand (for heavier elements) larger compared 
to the one for coherent diffraction of $12$ keV photons causing 
ultra-fast delocalization of core level (via the direct photo effect) 
and outer-shell electrons (via the Auger effect with typical time 
constants of below $10$ fs) culminating in the question whether 
the molecular structure will be first imaged and then destroyed 
or vice versa. Presently, common knowledge is that the maximum 
tolerable pulse length will be $10$ fs and many investigations 
concentrate on the destruction issue placing e.g. scarifying layers 
around the object of interest \cite{7, 8}.

As an alternative 'table-top' method, femtosecond electron 
diffraction (FED) has been suggested and developed to reach the 
above goals on various fronts using 'conventional guns' \cite{9,10}, 
envisioning intense laser accelerated electron bunches \cite{11} 
or using 'rescattered' electrons in above threshold ionization \cite{12}. 
Due to an elastic scattering cross section that is larger by about 
a factor of $\sim 10^6$ for $30$ keV electrons compared to $12$ 
keV photons along with the fact that inelastic, destructive reactions 
(electron impact excitation) occur with smaller cross sections and 
a factor of thousand less energy deposition compared to photon absorption, 
the number of electrons in the beam needed to record images as well 
as unwanted modifications of the sample molecules are significantly 
reduced \cite{10}. In the most promising 'conventional gun' scenario, 
bunches of up to $10^5$ electrons of $30$ keV within $\sim 600$ fs 
focused to spot sizes of $200$ $\mu$m have been demonstrated to be feasible 
recently and structural information has been achieved. $100$ fs pulse 
durations for bunches of $10^4$ electrons seem to be feasible optimizing 
existing gun designs \cite{10}. 

Both methods, X-ray and electron diffraction 
have been extensively discussed and compared in literature \cite{10, 13}. 
Drawbacks on the electron side are that it will be very challenging 
if not impossible to reach $100$ fs time scales and below, the relatively 
large beam divergence and, thus, lower spatial resolution, missing 
coherence and, most challenging, the so-called phase match problem 
when extended samples ($200$ $\mu$m) have to be used (like molecular 
ensembles in the gas phase): Due to the difference in velocity between 
the pump photons, initiating the dynamics, and the probe electron 
pulse of about a factor of four ($30$ keV electrons), the latter 
needing around picoseconds for traversing the sample region, 
a phase-mismatch between pump and probe occurs not to be overcome 
by shortening the pulses, essentially absent in all optical set-ups.

\begin{figure}[t] 
\vspace{-0.5cm}
\begin{center}
\includegraphics[width=0.47\textwidth]{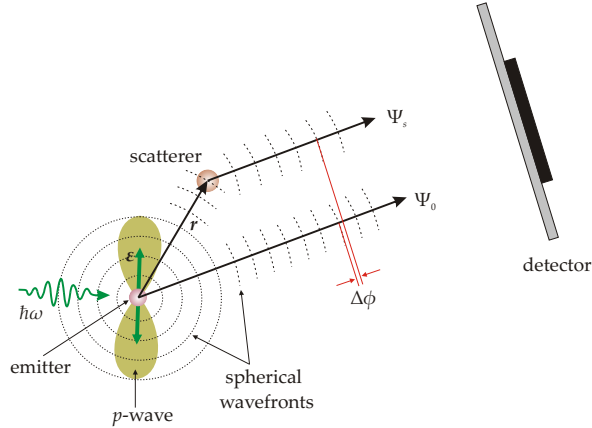}
\end{center} 
\vspace{-0.5cm} 
\caption{ The principle of photoelectron holography (as outlined e.g. in \cite{20}). 
A photoelectron p-wave is emitted from one of the atoms in the molecule 
via absorption of linear polarized light with photon 
energy $\hbar \omega_{\gamma}$ from e.g. the K-shell 
of the emitter with emerging spherical wavefronts. 
In photoelectron holography a photoelectron diffraction 
pattern is viewed as an interference between that part 
of the photoemitted wave which propagates undisturbed 
into the detector, $\Psi_0$, becoming a reference wave, 
and the part  scattered by the nearby atoms, $\Psi_s$, 
becoming an object wave which carries the information 
about the atomic positions. $\Psi_0$ and $\Psi_s$ 
interfere on the detector with a phase shift $\Delta \phi$ 
yielding thus an intensity distribution 
$I_{\bf p} = |\Psi_0+\Psi_s|^2$ 
which represents a hologram in a large portion 
of ${\bf p}$-space. }  
\label{figure1}  
\end{figure} 

In this paper we present a third scheme along with model calculations, 
combining the best of the above-described worlds enabling us to realize 
the molecular movie with femtosecond time and {\AA}ngstr\"om spatial 
resolution for small and medium sized molecules. It relies on the availability 
of X-ray FELs as well as of novel high-speed read-out, energy dispersive 
electron imaging devices (pnCCD see e.g. \cite{14}) and is based on the well-known 
photoelectron holography scenario (see e.g. \cite{15}-\cite{17} and references therein). 
Along those lines and as schematically depicted in figure \ref{figure1},  
it is suggested to utilize the X-ray FEL pulse 
to produce high-energy ($500$ eV to $2$ keV), femtosecond-pulsed and monochromatic 
photoelectrons in individual molecules in a gas-phase sample by exploiting 
the otherwise 'destructive' photo absorption effect with its large cross section. 
As illustrated in figure \ref{figure1}, emitted photoelectrons 
with energies specific to the atom(s) in the molecule that absorbed the photon(s) 
and, thus, launched at well-defined atomic sites, will on 'their way out' 
be diffracted on the potential of the (parent) molecule 
of interest with the diffraction pattern revealing, among others, the three 
dimensional relative Cartesian coordinates, i.e. the spatial structure of the molecule.

The method exploits the fact that photo absorption cross sections 
are of the order of $\sigma_{photo}\sim 10^{-20}$ cm$^2$ at about $1$ keV above an edge and, 
thus, typically four orders of magnitude larger than those for coherent 
photon scattering with $\sim 10^{-24}$ cm$^2$ for $10$ keV photons and 
carbon atoms. With a target line density of $\sim 10^{11}$ molecules/cm$^2$ 
and a photon flux of $10^{13}$ photons/pulse one could produce up to 
$10^4$ electrons/pulse. Since each individual electron is created 
'on the spot' with atomic spatial precision, i.e. at individual molecules, 
it represents a huge effective current density in the order of $10^{10}$ A/cm$^2$ 
per electron on an area of ($10$ {\AA})$^2$ for an electron velocity of several 
$10^9$ cm/s corresponding to keV energies and, thus, the total effective current density 
of $10^4$ electrons in the pulse reaches $\sim 10^{14}$ A/cm$^2$ per pulse 
and $\sim 10^{16}$ A/cm$^2$ per second for an FEL repetition rate of $120$ Hz. 
This compares with optimistic $\sim 10^8$ A/cm$^2$ per second that might 
be achieved in FED experiments assuming $\sim 10^4$ electrons/pulse focused 
to a spot diameter of $200$ $\mu$m with a pulse length of $100$ fs, an electron 
energy of $30$ keV and a gun repetition rate of $1$ MHz. 

Therefore, if combined with a large solid-angle, high-efficiency multi-hit, 
energy dispersive detectors, 
the method allows one exploring a whole class of molecules that can only 
be prepared as dilute samples with densities between $10^6$ to $10^{11}$ cm$^{-3}$ 
as typical for aligned \cite{18}, oriented \cite{19} 
or even conformer selected ensembles \cite{20, 21},  
all not accessible for pump-probe structure determination by any other present 
techniques. Here we built on the tremendous recent development in adiabatic \cite{18} 
or pulsed laser alignment (for a recent review see \cite{22}) as well as in Stark 
acceleration \cite{23} techniques to manipulate and define the molecular state 
where an unprecedented degree of laser induced alignment 
and orientation has been demonstrated \cite{19}.

The present paper is aimed to provide the basic ideas and numbers 
for femtosecond photoelectron diffraction (FPED), illustrated using 
the chlorine benzene molecule and verified by simple model calculations 
based on the well-known concepts of photoelectron holography. 
We also point to the limitations of the method and develop realistic 
scenarios for the three-dimensional momentum imaging of 
the photoelectrons by exploiting fore-front pnCCD X-ray pixel cameras.

\section*{Basic Concept and Comparison with other Methods} 

In the scenario illustrated in figure \ref{figure1}, we produce the photoelectron 
inside of an individual molecule in the sample with femtosecond time 
and $\sim 10^{-3}$ - $10^{-4}$ relative energy resolution, 
ultimately limited by the FEL radiation properties (here, seeded 
beams are envisioned for the future with $1$ fs or even sub-fs 
time-resolution). On its way out the photoelectron wave, 
launched at one specific atom of the molecule (emitter) 
is partly diffracted on the individual atoms of the parent 
molecule (scatterers), as depicted in figure \ref{figure1} 
for the most simple situation of only one scatterer, or reaches the detector 
directly, representing a reference wave for the scattered part. 
As described in \cite{16} the electron wave with wavelength $\lambda$ 
is fully coherent over a length of $l_c =  \lambda^2/\Delta \lambda$, 
i.e. over distances of $\sim 10$ nm to $\sim 0.1$ nm in the worst case, 
safely across medium-sized molecules, for an energy uncertainty 
of the FEL beam of $\Delta E_{\gamma}/E_{\gamma} \sim 10^{-4}$ at a few keV, 
translating in an uncertainly of the photoelectron energy of  
$\Delta E_e/E_e \sim 10^{-3}$-$10^{-4}$. Thus, a hologram is generated 
at the detector and, if the diffraction pattern is recorded over 
a large part of the solid angle, it allows one an immediate 
interpretation for photoelectron energies beyond $\sim 500$ eV. 

At the same time we transfer the nm wavelength of $1$ keV to $10$ keV 
photons to an to electron deBroglie wavelength $\lambda_e = h/\sqrt{2m_e E_e}$ 
between $\sim 0.6$ {\AA} to $0.3$ {\AA} for $E_e = 500$ eV or $E_e = 2$ keV 
photoelectrons, respectively, allowing us to achieve 
{\AA}ngstr\"om spatial resolution as demonstrated recently \cite{17}. 

In order to retrieve the structure of the molecule from the hologram 
on the detector three-dimensional electron momentum ${\bf P}_e=(p_x,p_y,p_z)$ images 
have to be recorded covering a large part of the final momentum space. 
This is achieved with an energy dispersive pixel pnCCD detector as described 
later, determining two momentum components 
from the electron hitting position on the detector and the third one 
from the measured energy of the photoelectron 
$E_e = \hbar \omega{_\gamma} - I_{n,l}$ via energy-momentum conservation relations  
($\hbar \omega{_\gamma}$ is the photon energy and $I_{n,l}$ is the binding energy 
of the atomic $n,l$ shell that is photo ionized).  
Electrons are emitted from molecules that are aligned, 
oriented or conformer selected by methods \cite{18}-\cite{23} 
described in some detail later. 

The method as proposed here comprises several new concepts, 
builds on fore-front technologies in various areas and combines 
them in a unique way:   

(i) Exploiting the new FEL light sources delivering intense 
($10^{13}$ photons per pulse), short-time ($\sim 100$ fs with 
prospects of achieving $1$ fs or even below) 
and coherent VUV or X-ray pulses with energies of 
$100$ eV $< \hbar \omega_{\gamma} < 12$ keV \cite{1}-\cite{5}.  

(ii) Building on the ability to align, orient or conformer select 
in three dimensional space the molecules of interest with 
major progress achieved in the recent past \cite{18}-\cite{23}.  

(iii) Using high-energy electrons ($500$ eV $< E_e < 2$ keV) 
for diffraction where the patterns are much simpler for 
interpretation in terms of photoelectron holography 
\cite{15}-\cite{17}, \cite{23a,23b}.  

(iv) Applying a new concept for $3D$ electron momentum 
imaging based on recently developed, large area, energy and 
position dispersive, single electron counting pnCCD detector devices \cite{14}.  

(v) Creating the photoelectrons directly by the X-ray pulse 
we do have an effective 'all optical approach' in pump-probe experiments, 
thus, not facing the phase matching problem as in FED measurements 
using conventional electron guns \cite{10}. 

Thus, our method represents a major step forward compared 
to all previous concepts as e.g. so called molecular frame 
photoelectron angular distribution (MFPAD) spectroscopy being 
now a standard technology at synchrotron radiation facilities. 
Based on a recent experimental break-through in angle-resolved 
photoelectron-photoion coincidence techniques \cite{24}-\cite{26} culminating 
in using reaction microscopes (REMI) \cite{27}-\cite{33} that allow one for the coincident 
detection of the $3D$ vector momenta of several electrons and ions, 
low-energy photoelectron angular distributions from free, 'fixed-in-space' 
molecules are recorded. Here, in present experiments, the molecular 
orientation is retrieved a posteriori via coincident detection of heavy 
molecular fragments making this method extremely demanding technologically 
and limiting it to essentially diatomic molecules where, in addition,  
the axial recoil approximation has to hold. Since (multi-) coincidences 
are requested, typically less then one event per beam pulse can be accepted 
for a reliable analysis of the data, thus requiring MHz repetition rates 
from the pulsed radiation source as available at synchrotrons in order 
to take multi-dimensional data with sufficient statistical significance. 
Moreover, due to the timing properties of synchrotrons in the range of $100$ ps, 
no short-time pump-probe experiments can be realized. Femtosecond slicing 
devises available at some synchrotrons do not deliver enough photons 
for such investigations \cite{34}. Finally, only low-energy electrons 
$E_e < 50$ eV have been recorded up to now (manly due to the time-of flight 
spectroscopy methods used) that makes the interpretation of the data demanding. 
Nevertheless, scattering patterns provided detailed information to an unprecedented 
level about a variety of processes such as localization of charges \cite{30} 
and core holes \cite{31}, interferences in molecular double-slit 
or multi-slit arrangements \cite{32,35}, phases of photoelectron waves \cite{36,37}, 
and, last but not least, the electronic potentials and the molecular 
structure via photoelectron diffraction \cite{33,38}.

It also decisively extends the huge area of previous ultra-fast pump-probe, 'femto-chemistry' 
studies (see e.g. \cite{39}) performed with long-wavelength (optical) probe lasers 
that can only indirectly deliver final-state (time of probe) structure information 
of smaller molecules based on spectroscopic knowledge of the potential curves 
accessed by the probe pulse. Since the potential curves along 
the reaction coordinate usually cannot be calculated 
for larger molecules, this method remained limited to smaller species. 

Reaching femtosecond timescales without the need of crystallization 
of samples we can access a different class of molecules 
as well as ultra-short times, well below the picosecond 
time-resolved experiments recently performed at synchrotrons (see e.g. \cite{40}).

Last but not least our method is expected to have certain advantages 
compared to using VUV high-harmonic radiation as a 
probe pulse and which has already lead to first successful 
measurements (see e.g. \cite{41}).  In the latter, up to now, 
photons are not high-energetic enough at acceptable intensities 
to access core-level electrons and, thus, 
to specify in details the birth location of the photoelectron. 
Since in principle, attosecond time resolution can be achieved 
with this technology both methods might be considered as complementary. 

\section*{Theoretical Description and Results} 

Let us consider the photo ionization of a molecule 
assuming that this process can be regarded as an 
effectively single-electron problem. 
The corresponding Schr\"odinger equation reads  
\begin{eqnarray} 
i \frac{\partial \Psi}{\partial t} = %
\left(\hat{H}_0 + \hat{W}_{EM}(t)\right)\Psi.   
\label{schr}
\end{eqnarray}
Here, $\Psi=\Psi({\bf r},t)$ is 
the electron wave function which is 
space and time dependent and 
describes the dynamics of the 'active' electron 
in the photo ionization process. 
Further, $\hat{H}_0$ is the Hamiltonian for 
the electron in the absence 
of the electromagnetic field and  
\begin{eqnarray} 
\hat{W}_{EM}(t)={\bf A} \cdot \hat{\bf p}/c   
\label{em_interaction}
\end{eqnarray}
is the interaction between the electron and 
the field, where $\hat{\bf p}$ is the operator 
for the electron momentum, ${\bf A}$ is the vector potential 
of the field and $c$ is the speed of light. 
Assuming that the field is linearly 
polarized we take  
\begin{eqnarray} 
{\bf A}({\bf r},t) = {\bf A}_0 %
\cos(\omega_0 t - {\bf k}_0 \cdot {\bf r}).  
\label{laser-field}  
\end{eqnarray} 
Here ${\bf r}$ and $t$ are the space and time coordinates, 
$\omega_0$ and ${\bf k}_0$ are the frequency and wave vector 
and ${\bf A}_0 = a_0 {\bf e}$, where ${\bf e}$ 
is the polarization vector (${\bf e} \cdot {\bf k}_0 =0$) and 
$a_0 = c F_0/\omega_0$ is 
the amplitude of the vector potential with 
$F_0$ being the strength of the electromagnetic field.   

Below we shall discuss two approaches 
to address photo ionization.  
In one of them the process is dealt with by 
finding the wave function, which is a solution 
of Eq.(\ref{schr}) satisfying appropriate initial 
and boundary conditions 
and which describes the evolution of 
the electron wave packet in time and space. 
The second approach, which does not consider the space-time 
characteristics of the process, 
is based on obtaining the transition amplitude  
which enables one to calculate the momentum spectrum 
of the emitted photoelectrons.     
Clearly, these two approaches can be considered as complementary.  

\subsection*{Approach 1}

The wave function $\Psi=\Psi({\bf r},t)$ 
can be expanded according to 
\begin{eqnarray}
\Psi(t) &=& g(t) \psi_i(t) + 
\int d^3 {\bf p} \beta_{\bf p}(t) 
\psi_{\bf p}^{(+)}(t),   
\label{w1} 
\end{eqnarray} 
where $\psi_i(t)$ 
and $\psi_{\bf p}^{(+)}(t)$ are solutions of 
the field-free Hamiltonian 
\begin{eqnarray} 
i \frac{\partial \psi}{\partial t} = %
\hat{H}_0 \psi,    
\label{hamilton_0}
\end{eqnarray}
for bound and continuum states of the electron, 
respectively, and $g(t)$ and $\beta_{\bf p}(t)$ 
are the time-dependent expansion coefficients.  
  
Since in the (tightly) bound state  
the electron is very well localized we can 
approximate $\psi_i(t)$ as 
\begin{eqnarray}
\psi_i(t) &=& \varphi_i({\bf r} - {\bf R}_0) %
\exp(-i \varepsilon_0  t),  %
\label{e2} 
\end{eqnarray}  
where $\varphi_i$ is the state of the electron 
bound in the atom, whose nucleus 
has the coordinates ${\bf R}_0$,  
and $\varepsilon_0 $ 
is the corresponding binding energy. 
  
In order to build states $\psi_{\bf p}^{(+)}(t)$ 
representing the stationary continuum spectrum  
we shall assume that the kinetic energy 
of the electron motion in the continuum 
is sufficiently high and, therefore, $\psi_{\bf p}^{(+)}(t)$ 
can be obtained in the first order approximation 
in the interaction between the electron and 
the atomic centers constituting the molecule. 
This yields  
\begin{eqnarray}
\psi_{\bf p}^{(+)}(t) &=& %
\frac{\exp(-i E_p  t)}{(2 \pi)^{3/2}} \times %
\nonumber \\ 
&& \left( \exp(i {\bf p} \cdot {\bf r} ) + %
\frac{ 1 }{(2 \pi)^{3/2}} \sum_{j} %
\exp(i {\bf p} \cdot {\bf R}_j) \int d^3 {\bf p}' %
\frac{ \widetilde{ V }_j({\bf p}' - {\bf p}) }{E_p - E_{p'} + i \alpha } %
\exp(i {\bf p}' \cdot ({\bf r} - {\bf R}_j)) \right).  
\label{e3} 
\end{eqnarray}  
Here ${\bf p}$ and $E_p = p^2/2m_e$ ($p=|{\bf p}|$) are 
the momentum and energy, respectively, of 
the continuum state and  
\begin{eqnarray}
\widetilde{ V }_j({\bf q})= \frac{1}{(2 \pi)^{3/2}} 
\int d^3 {\bf r} V_j({\bf r}) %
\exp(- i {\bf q} \cdot {\bf r})
\label{e4} 
\end{eqnarray}  
is the Fourier transform of the interaction 
$V_j({\bf r})=V_j({\bf r}-{\bf R}_j)$ 
between the electron and the $j$-th atomic center 
of the molecule whose nucleus is located 
at a point with the coordinates ${\bf R}_j$.       
In Eq.(\ref{e3}) $ \alpha \to +0$ shows 
how to handle the singularity and  
the sum runs over all atomic centers 
of the molecule, $j=0, 1, ..., N-1$ where 
$N$ is the number of atoms,   
including the emitting one ($j=0$). 
One should note that in (\ref{e3}) we have 
neglected all inelastic channels 
corresponding to energy transfers between 
the emitted photoelectron and the internal 
degrees of freedom of the residual molecular ion.   
Note also that the state (\ref{e3}) is  
the so called 'out-state',  
which asymptotically (at large distances between 
the electron and the residual molecule)  
is a superposition of a plane and an 
outgoing scattered wave.  

In order to obtain the unknown coefficients  
$g(t)$ and $\beta_{\bf p}(t)$ we 
insert the wave function (\ref{w1}) 
into the Schr\"odinger equation (\ref{schr}) 
and use the rotating-wave approximation. 
This leads to the following system 
of differential equations 
\begin{eqnarray}
i \frac{d g(t) }{dt} &=& \int d^3 {\bf p} W_{g,{\bf p}} %
\exp(i \omega_0 t) \beta_{\bf p}(t) 
\nonumber \\ 
i \frac{d \beta_{\bf p}(t) }{dt} &=& %
W^*_{g,{\bf p}} \exp(i \omega_0 t) g(t),   
\label{w2} 
\end{eqnarray} 
where 
\begin{eqnarray}
W_{g,{\bf p}} = \frac{1}{2c} \int d^3 {\bf r} 
\psi^*_i(t=0) \exp(- i {\bf k} \cdot {\bf r}) %
({\bf A}_0 \cdot \hat{\bf p}) \psi_{\bf p}^{(+)}(t=0). 
\label{w3} 
\end{eqnarray} 
Assuming for the moment that the electromagnetic field 
is suddenly switched on at $t=0$, when the electron 
was in the state $\psi_i$, the initial conditions 
for the system (\ref{w2}) are given by  
$g(t=0)=1$ and $\beta_{\bf p}(t=0)=0$.  

Here we shall not go in details of how the system 
(\ref{w2}) can be solved and the state (\ref{w1}) derived 
and merely note that one can show that at asymptotically 
large distances between the photoelectron and the residual molecule 
the wave function $\Psi(t) $ can be presented 
in the following approximate form   
\begin{eqnarray}
\Psi({\bf r},t) &=& \exp(-i E_e t) \times 
\left( \frac{ \exp( i p_e \xi_0) }{ \xi_0 } 
\exp(-\Gamma (t- \xi_0/v_e))
\theta(t- \xi_0/v_e) \right. 
\nonumber \\  
&& \left. \times \sum_{l=0}^{\infty} \sum_{m=-l}^{+l} 
B(l_i,m_i;l,m;\mbox{\boldmath$\xi$}_0/\xi_0) 
+ \right. 
\nonumber \\ 
&& \left. 
\sum_{j\neq 0}  
\frac{ \exp( i p_e \xi_j ) }{ \xi_j } 
\frac{ \exp( i p_e R_{j0}) }{ R_{j0} }
\exp(-\Gamma (t- (\xi_j+R_{j0})/v_e))
\theta(t- (\xi_j+R_{j0})/v_e) \right.  
\nonumber \\ 
&& \left. \times  %
\sum_{L=0}^{\infty} \sum_{M=-L}^{+L}\sum_{l=0}^{\infty} \sum_{m=-l}^{+l} 
C(l_i,m_i;L,M;l,m; \mbox{\boldmath $\xi$}_j/\xi_j, {\bf R}_{j0}/R_{j0}) \right).  
\label{w4} 
\end{eqnarray} 
Here, $E_e = \varepsilon_0 + \omega_0$ is the energy of the emitted electron, 
$\Gamma$ ($\Gamma \ll E_e$) is the half-width of the initial electron 
state caused by the photo effect, 
$p_e = \sqrt{2 m_e E_e}$ and $v_e = p_e/m_e$ are the absolute 
values of the electron momentum and velocity, respectively,    
$\mbox{\boldmath $\xi$}_j = {\bf r} - {\bf R}_j$, 
${\bf R}_{j0} = {\bf R}_j - {\bf R}_0$,   
$l_i$ and $m_i$ are the angular momentum and its projection 
in the initial electron state $\psi_i$ and $\theta$ 
is the (stepwise) theta-function. 
The functions $B(l_i,m_i;l,m;\mbox{\boldmath$\xi$}_0/\xi_0)$ and 
$C(l_i,m_i;L,M;l,m; \mbox{\boldmath $\xi$}_j/\xi_j, {\bf R}_{j0}/R_{j0})$  
have rather complicated forms  
and will be specified elsewhere. 

The physical meaning of the state Eq.(\ref{w4}) 
is rather transparent. This state describes an electron wave packet 
which is a superposition of the wave, propagating directly 
from the initial electron location 
(the first two lines in the parentheses), 
and of the 'secondary' waves, appearing due to the electron scattering 
on all other atomic centers in the molecule (the last two lines).    
The theta-functions in Eq.(\ref{w4}) emphasize 
the fact that in order 
to traverse a distance $L$ an electron 
moving with a velocity $v_e$ needs the time $L/v_e$ 
and the decaying exponential factors  
reflect the depletion of the electron population 
in the initial state due to the emission.  

\subsection*{Approach 2} 

The spectrum of photoelectrons emitted 
from the molecule can be calculated 
using the transition amplitude  
\begin{eqnarray}
S_{fi} = -i \int_{-\infty}^{+\infty} dt %
\langle \psi_f(t) |\hat{W}_{EM} |\psi_i(t)\rangle.       %
\label{trans_ampl} 
\end{eqnarray}  
where $ \psi_i(t)$ is given by Eq.(\ref{e2}) and 
$\psi_f(t)$ is the so called 'in-state' for 
the emitted electron, $\psi_f(t)= \psi_{\bf p}^{(-)}(t) $, 
which is obtained from the 'out-state' (\ref{e3}) 
according to $\psi_{\bf p}^{(-)}(t) = 
\left( \psi_{-\bf p}^{(+)}(-t)\right)^*$. 
Inserting the states $ \psi_i(t)$ and 
$\psi_{\bf p}^{(-)}(t)$ into (\ref{trans_ampl})
and integrating over the time and space coordinates 
we obtain 
\begin{eqnarray}
S_{fi} &=& -\frac{ \pi i }{c} \exp(i {\bf k} \cdot {\bf R}_0) %
G_{fi} \times \delta\left(E_p - \varepsilon_0 - \omega_0\right),  
\label{e5} 
\end{eqnarray}   
where 
\begin{eqnarray}
G_{fi} &=& ({\bf A}_0 \cdot {\bf p}) %
\widetilde{ \varphi }_i({\bf p}-{\bf k}) %
\exp( -i {\bf p} \cdot {\bf R}_0 ) %
+ \frac{ 1 }{ (2 \pi)^{3/2} } \sum_{j} %
\exp( - i {\bf p} \cdot {\bf R}_j ) %
\nonumber \\ 
&& \times \int d^3 {\bf p}' %
\frac{ \widetilde{ V }_j({\bf p} - {\bf p}') }{ E_p - E_{p'} + i \alpha } %
\exp(i {\bf p}' \cdot ({\bf R}_j - {\bf R}_0)) %
({\bf A}_0 \cdot {\bf p}') \widetilde{ \varphi }_i({\bf p}'-{\bf k}).    
\label{e6} 
\end{eqnarray}   
The $\delta$-function in (\ref{e5}) expresses 
the energy conservation in the photo effect and 
\begin{eqnarray}
\widetilde{ \varphi }({\bf q})= \frac{1}{(2 \pi)^{3/2}} 
\int d^3 {\bf r} \varphi({\bf r}) %
\exp(- i {\bf q} \cdot {\bf r})
\label{e7} 
\end{eqnarray}  
is the Fourier transform of the initial electron state. 

Using the standard consideration for finding 
the momentum distribution of the photoelectrons 
from the transition amplitude we obtain 
that this distribution is given by  
\begin{eqnarray}
\frac{ d \sigma}{d^3 {\bf p}} &=& \frac{ \pi  }{2 c^2} %
|G_{fi}|^2 \times \delta\left(E_p - \varepsilon_0 - \omega_0\right).   
\label{e8} 
\end{eqnarray}   
The knowledge of the momentum distribution of 
the emitted electrons and the experimental geometry 
enables one to calculate the electron intensity 
on the detector. 

\subsection*{Results}   

In the present paper we will not present results 
obtained with the wave function (\ref{w4}). 
Instead, assuming that $\Gamma \to 0$ 
and considering only the space-time points 
with $v_e t > \xi_j+R_{j0} $,   
we shall replace expression (\ref{w4}) 
by a simpler one which was earlier used 
in the studies of inside photoelectron 
holographic imaging of solids \cite{15}-\cite{17}. 

\begin{figure}[t] 
\vspace{-0.5cm}
\begin{center}
\includegraphics[width=0.47\textwidth]{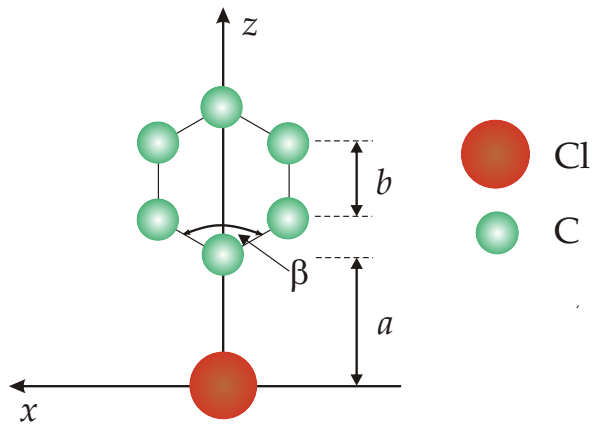}
\end{center} 
\vspace{-0.5cm} 
\caption{ A sketch of the chlorine benzene molecule  
(the hydrogen atoms are not shown):  
$a=2.01$ {\AA}, $b=1.4$ {\AA} and 
$\beta = 2 \alpha =120^\circ$.}  
\label{figure2}  
\end{figure} 

\begin{figure}[t] 
\vspace{-0.5cm}
\begin{center}
\includegraphics[width=0.33\textwidth]{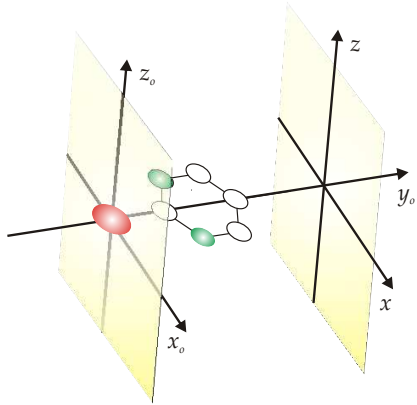}
\end{center} 
\vspace{-0.5cm} 
\caption{ A sketch of geometry I 
showing the emitter (Cl atom) 
which is taken as the origin of the coordinate system and 
two 'active' scatterers (C atoms) located 
in the $x_{0}$-$y_{0}$ plane. }   
\label{figure3a}  
\end{figure}

We take the principle scenario depicted 
in figure \ref{figure1} and consider 
the photo ionization of a chlorine molecule 
(see figure \ref{figure2}) from the $K$-shell of 
the Cl atom by a beam of $4.5$ keV FEL photons. 
The photons are assumed to be linearly polarized 
along the $y$-axis in the case of geometry I 
(figure \ref{figure3a}) and along the $z$-axis in the case 
of geometry II (figure \ref{figure6}).
The photon absorption results in the emission of a $1.7$ keV   
electron which scatters on the carbon atoms of the molecule   
before reaching the detector (we neglect 
the hydrogen atoms as scatterers because 
of their relatively weak field). 

A monochromatic electron source $p$-wave $\Psi_{source}$ 
is launched from the $K$-shell of Cl by dipole transition 
\begin{eqnarray} 
\Psi_{source}= \frac{A \exp(i p_e \xi_0 )}{\xi_0} %
Q_{1}(\xi_0) Y_{1,0}(\mbox{\boldmath$\xi$}_0/\xi_0)   
\label{s-model-2}
\end{eqnarray} 
with the radial part $Q_{1}$ and spherical 
harmonic $Y_{1,0}$ describing the electron preferably ejected along 
the electric field vector.  

In the single scattering picture, the total 
wave $\Psi_{tot}$ on the detector plane is 
\begin{eqnarray} 
\Psi_{tot} = \Psi_{source}(\mbox{\boldmath$\xi$}_0)    
+ \sum_j f_j(\Theta_j) \frac{ \exp(i p_e \xi_j) }{ \xi_j } %
\Psi_{source}(\mbox{\boldmath$\xi$}_0 - \mbox{\boldmath$\xi$}_j).   
\label{s-model-3}
\end{eqnarray} 
Here $\Theta_j$ is the scattering angle and  
$f_j(\Theta_j)$ the scattering amplitude, 
which in the first Born approximation is given by   
\begin{eqnarray} 
f_j(\Theta)= - \frac{m_e}{2 \pi \hbar^2} 
\int d^3 \xi V_j(\mbox{\boldmath$\xi$}) 
\exp(-i {\bf q} \cdot \mbox{\boldmath$\xi$}),      
\label{s-model-4}
\end{eqnarray} 
where $|{\bf q}| = 2 p_e \sin(\Theta/2) $. 
The simple expression (\ref{s-model-3}) should be 
a reasonable approximation if the size of the space, 
where the scattering potential $V_j$ is effectively located, 
is substantially less than the distance $R_{j0}$. 
The Born approximation is  
safely valid for $\sim 1$ keV photoelectrons. 

\begin{figure}[t] 
\vspace{0.25cm}
\begin{center}
\includegraphics[width=0.47\textwidth]{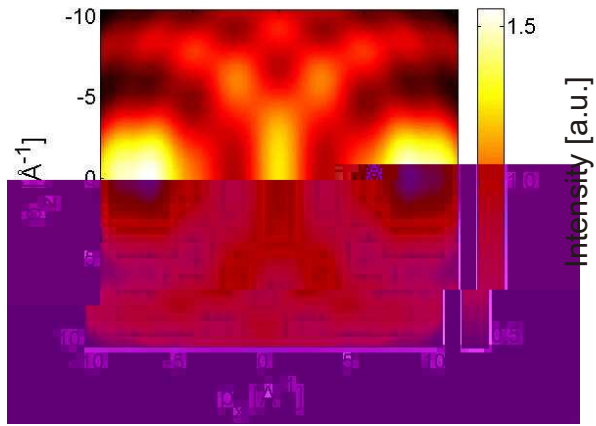}
\end{center} 
\vspace{-0.5cm} 
\caption{ The calculated photoelectron 
hologram for the geometry I. 
Here, a $1.7$ keV photoelectron 
emitted from the Cl atom is scattered by two C atoms.}   
\label{figure3b}  
\end{figure} 

\begin{figure}[t] 
\vspace{-0.5cm}
\begin{center}
\includegraphics[width=0.47\textwidth]{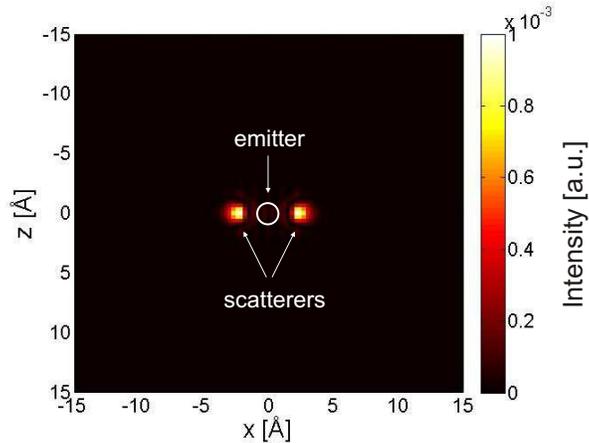}
\end{center} 
\vspace{-0.5cm} 
\caption{ Atomic images in the plane 
$x$-$z$, $6$ {\AA} above the emitter, 
reconstructed from the hologram shown in in figure \ref{figure3b}. 
As usual with inside-source holography, 
the emitter (indicated by a circle) sits 
at the origin of the image and is not reproduced.}   
\label{figure3c}  
\end{figure}  

In a first step we take geometry I sketched in 
figure \ref{figure3a} and   
only consider two out of the six carbon atoms 
of the molecule which are marked 
in figure \ref{figure3a} by solid circles.  
The calculated hologram $I(p_{x},p_{z})=|\Psi_{tot}|^{2}$ 
is shown in figure \ref{figure3a}. The hologram $I(p_{x},p_{z})$ can be inverted 
by mathematical means to yield real space images that locate individual 
atoms surrounding the emitter. Here we adopt a reconstruction formalism 
based on the Helmholtz-Kirchoff integral theorem as proposed 
in \cite{41a}, where the amplitude of the object wave field 
at any point $\textbf{r}$ in space near the emitter can be calculated 
by Fourier transforming phased two-dimensional photoelectron hologram.  
In principle it should be possible to achieve a sub-{\AA}ngstr\"om spatial 
resolution $\delta r\propto \pi / p_{e} $, however  the anisotropy 
of the electron scattering (with forward scattering being a major obstacle) 
produces artifacts in the reconstructed images which are manifested 
as a small distortions and peak shifts from the real space 
images amounting about $1$ {\AA}. Several procedures 
to correct these effects and thus to improve the image 
quality and/or extract additional informations have 
been proposed: some of them are based on the inversion 
formalism \cite{41b}, \cite{17} whereas others exploit 
small changes in diffraction condition 
(see, for example, \cite{23a}). 
\begin{figure}[t] 
\vspace{0.25cm}
\begin{center}
\includegraphics[width=0.33\textwidth]{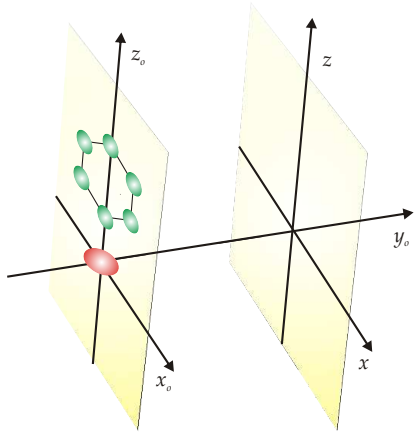}
\end{center} 
\vspace{-0.5cm} 
\caption{ A sketch of geometry II showing 
the emitter (Cl atom) which is taken as 
the origin of the coordinate system 
and six scatterers (C atoms) 
on the $x_{0}-z_{0}$ plane. }   
\label{figure4a}  
\end{figure} 
 
\begin{figure}[t] 
\vspace{-0.5cm}
\begin{center}
\includegraphics[width=.47\textwidth]{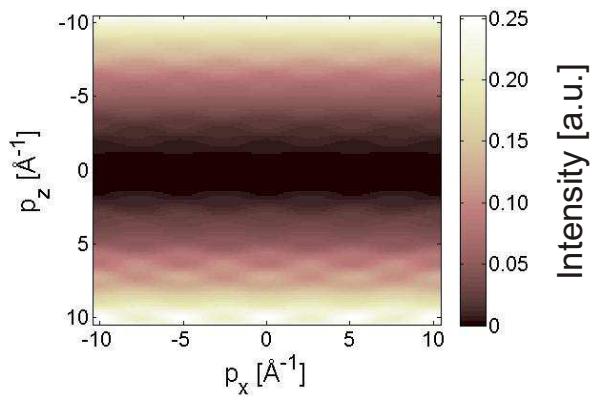}
\end{center} 
\vspace{-0.5cm} 
\caption{ The calculated photoelectron hologram 
for geometry II. Here a $1.7$ keV photoelectron 
emitted from the Cl atom is scattered 
by six C atoms located on the $x_{0}$-$z_{0}$ plane. }   
\label{figure4b}  
\end{figure} 

\begin{figure}[t] 
\vspace{0.25cm}
\begin{center}
\includegraphics[width=0.47\textwidth]{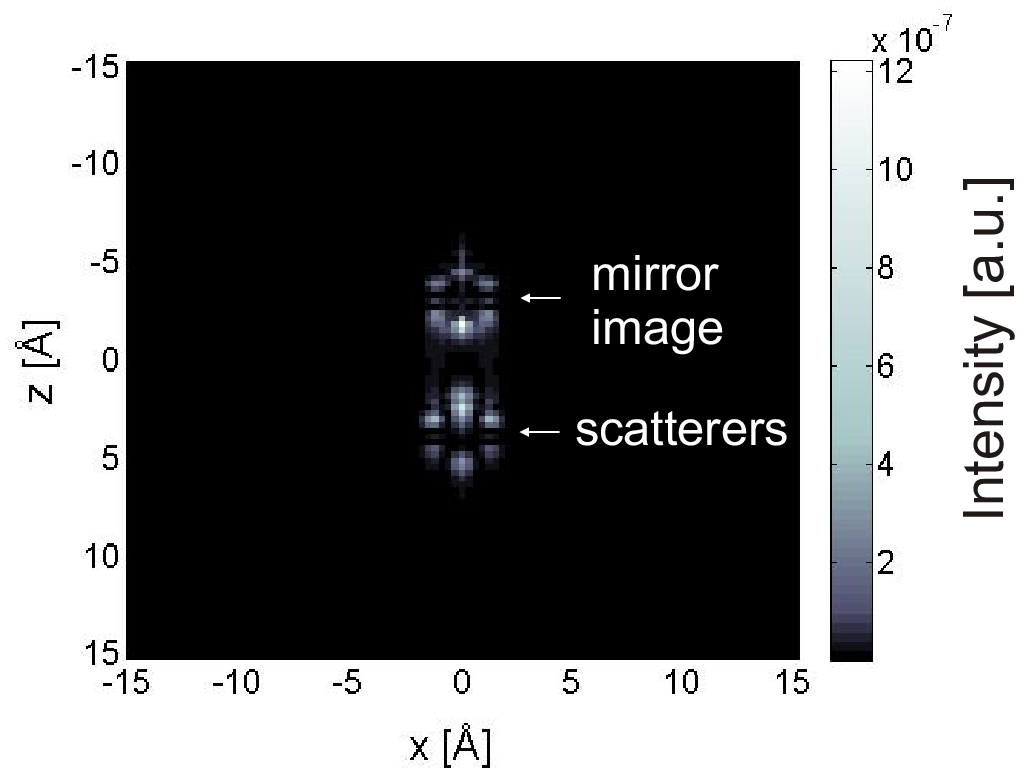}
\end{center} 
\vspace{-0.5cm} 
\caption{ Positions of the scatterers 
(and the mirror image) $1$ {\AA} 
above the emitter retrieved from 
the hologram shown in \ref{figure4b}.}   
\label{figure4c}  
\end{figure}  

Next we assume geometry II sketched in figure \ref{figure4a} 
and consider now all six carbon atoms as scatterers.   
In this case we obtain an interference hologram shown 
in figure \ref{figure4b}. 
Retrieval of the structure yields 
several spots at the positions 
of the carbon nuclei as well 
as their mirror images (see figure \ref{figure4c}). 

\begin{figure}[t] 
\vspace{-0.5cm}
\begin{center}
\includegraphics[width=.37\textwidth]{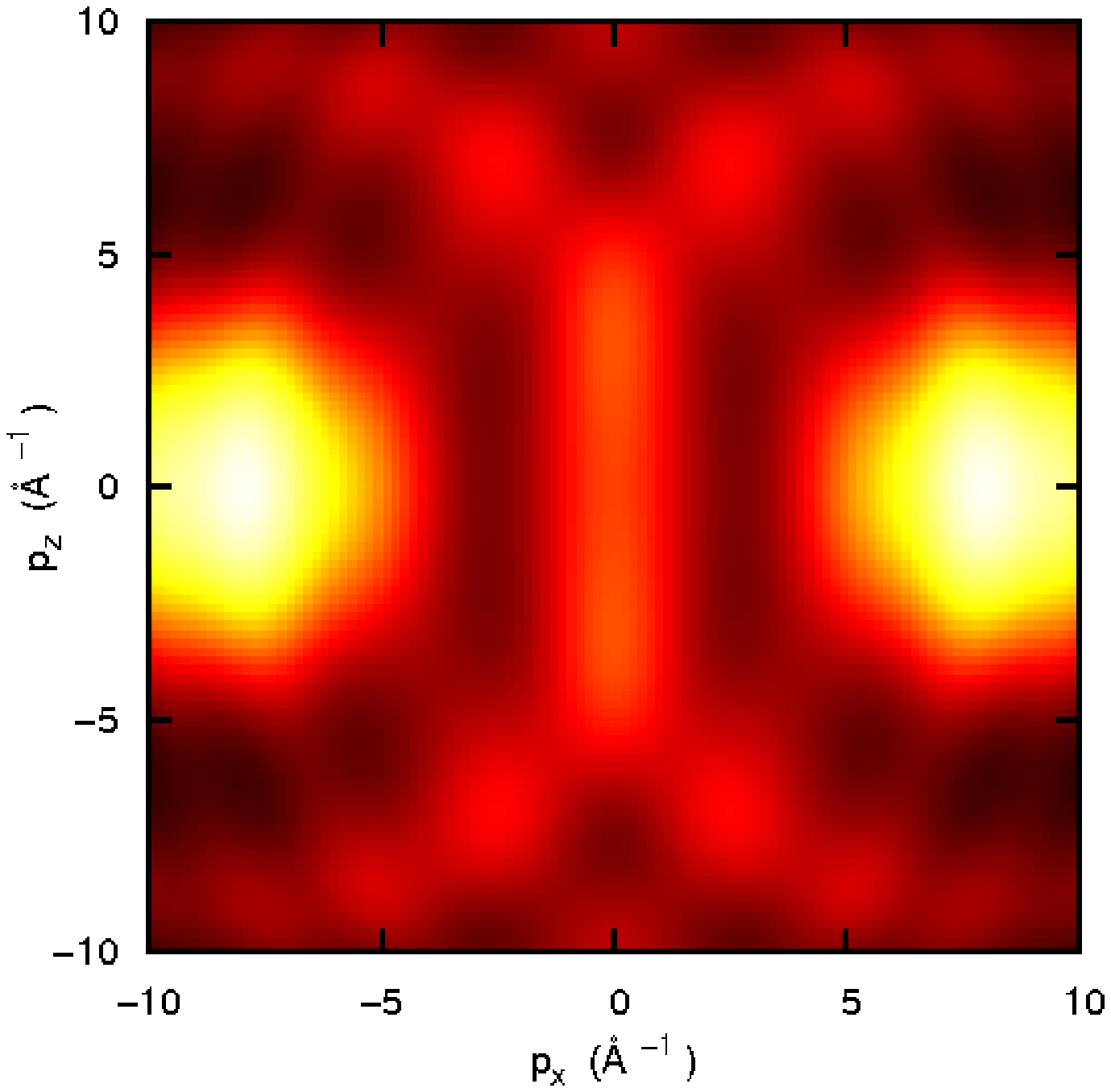}
\end{center} 
\vspace{-0.5cm} 
\caption{ The case of Geometry I.  The hologram 
on the detector calculated using the approach 2.}    
\label{figure3bennaceur}  
\end{figure} 

Note also that we have performed calculations 
for the system and geometries using the approach 2. 
We have found that, compared to the description discussed above, 
these calculations lead to holograms, 
which possess the same qualitative features 
(see for instance figure \ref{figure3bennaceur})  
and, thus, yield the same molecular structure.  

Summarizing the above brief discussion we can conclude that 
our quantum calculations, that should be sufficiently accurate 
at this high photon energy, reveal the salient interference features 
due to the scattering of the outgoing photoelectron 
from the different atoms in the molecule 
and, thus, support the basic idea brought forward in this paper. 

\section*{ Possible technical realization using chlorine benzene 
as a model system} 

{\bf Radiation source}: LCLS provides $N_{\gamma}\sim 10^{13}$ photons 
at up to $10$ keV in $75$ fs pulses at a design repetition rate of $120$ Hz which 
will be the basis of our estimates. We assume to focus the radiation 
to a spot diameter of $10$ $\mu$m. Recently, lasing at $8.3$ keV has been 
demonstrated with a $10^{12}$ photons per pulse 
at a width of $75$ fs and beam-time will be allocated 
in late 2010 at the XPP endstation \cite{4}. 

\vspace{0.25cm} 

{\bf Aligned target}: Rotationally cold molecules can be delivered 
by a supersonic jet using a pulsed Even-Lavie nozzle \cite{42,43}. 
Intersecting the cold molecular jet in the interaction region 
with pulses from a seeded, nanosecond Nd:YAG alignment laser, 
synchronized to the LCLS pulses yields adiabatic $1D$ and $2D$ alignment 
of the molecular samples in space. Orientation of molecules in $1D$ or $2D$, 
the latter meaning that the molecule is controlled in all three spatial 
dimensions can be achieved using mixed laser and dc electric fields. 
Adopting the numbers put forward in one of the proposals to LCLS on coherent 
diffraction imaging on adiabatically aligned dibromo benzene, 
densities up to $10^{11}$ cm$^{-3}$ $2D$ aligned chlorine benzene molecules can 
be achieved resulting in a target line density of $\Delta x=10^{10}$ cm$^{-2}$ 
assuming a target that is $0.1$ cm extended along the FEL beam direction 
which will be the basis of the present estimates. Applying additional 
fields for orienting the molecules or even obtaining conformer selected 
samples significantly decreases the target densities to values that 
might be as small as $10^6$ cm$^{-3}$. 

\vspace{0.25cm} 

{\bf Photoelectron energies and absorption cross sections}: 
Photon absorption can take place in any electronic state of the 
molecule with strongly varying cross sections. Large values 
are achieved close to edges, where the photon energy matches 
the binding energy of the respective electron resulting 
in typical cross sections of $\sim 10^{-18}$ cm$^2$ and producing 
low-energy photoelectrons. Above the edge, the cross section for 
that specific shell decreases rapidly with about $(\hbar \omega_{\gamma})^{-3.5}$. 
Assuming that core-level electrons of the atomic constituents of the molecule 
are essentially characterized by their atomic energy levels and cross sections 
we adopt one situation in which $4.5$ keV photons irradiate the molecules. 
In such a case the dominant channel of the electron emission is 
photon absorption by the $K$-shell electrons of Cl with the cross section 
of $ \sim 10^{-20}$ cm$^2$.  
The other absorption channels are not only much weaker but also 
lead to the emission of electrons having much higher energies. 
Therefore, they can be easily separated from the main channel 
and in what follows will simply be ignored. 

\vspace{0.25cm}

{\bf Numbers of emitted photoelectrons}: 
Assuming an electron detection efficiency of one we can estimate 
the number of recorded photoelectrons per pulse according 
to $N_{pe} = \Delta x \sigma_{photo} N_{\gamma}$ 
ending up with about $N_{pe}$(Cl$_K$) $\sim 10^3$. 
With a repetition rate of 120 Hz the corresponding rate  
would be $1.2 \times 10^{5}$ electrons per seconds. 
In order to estimate the time required to record a snapshot of 
the molecular structure one needs in addition the elastic scattering 
cross section for the interaction of the launched photoelectrons with 
the atoms in the molecule on their way out which is implicitly included 
in our later theoretical simulations. 
Taking into account the p-wave character of the emitted photoelectrons 
and assuming that each of them interacts with its parent molecule only, 
the count rate, which in addition depends on the relative orientation 
of the molecule with respect to the (dipolar) photoelectron emission 
characteristics and the position of the detector, will be between $5000$ 
electrons/s and $60000$ electrons/s. Assuming $300000$ electrons 
to be detected in order to obtain structural information of a medium 
sized molecule at a certain time delay between pump and probe pulses, 
a full image might be recorded in a matter of seconds 
to few minutes for one time step.

From these considerations it is obvious that ultra-dilute samples 
with line densities as small as $10^6$ cm$^{-2}$ like for conformer selected 
molecules only become accessible at photon energies slightly above 
one of the edges of the atomic constituents of the molecule 
exhibiting cross sections on the order of $10^{-18}$ cm$^2$ and photoelectron 
energies of not more than $100$ eV. Nevertheless, assuming the full 
performance of the LCLS with repetition rates as high as $120$ Hz 
we still end up with about $10$-$120$ photoelectrons per second such 
that a measurement of the structure, again requesting about $300 000$ events 
would take about two to ten hours.  

From these considerations it becomes also obvious that direct 
coherent X-ray diffraction of oriented or even conformer selected 
molecules with elastic photon scattering cross sections 
of the order of $10^{-24}$ cm$^2$ are completely unrealistic! 
The same holds for FED where in the most 
optimistic case $10^4$ electrons in $100$ fs pulses might hit the target 
at $1$ MHz repetition rate. With elastic scattering cross sections 
on the order of $10^{-19}$ cm$^2$ and a target line density of 
$10^6$ molecules/cm$^2$ this yields a few scattered electrons per hour. 

Thus, the technology presented here does indeed allow one to access  
a large number of samples that are of fundamental and benchmarking 
interest for tracing molecular dynamics during chemical reactions. 
For lower photoelectron energies, not considered in this article 
but as used presently at synchrotrons in similar, time-independent 
experiments the speed for taking structural images would be limited 
by the count rates accepted by the various detectors as well as 
by space charge effects blurring 
the image forcing one to reduce the incoming flux.

\vspace{0.25cm}
 
{\bf Energy Dispersive Large-Area Electron Momentum Imaging. } 

The new method for obtaining structural information of atomic positions 
with {\AA}ngstr\"om resolution in molecules as a function of time 
via femtosecond holographic imaging of photoelectrons relies, 
apart from the availability of X-ray FELs and of $2D$ oriented molecular 
samples, decisively on high-speed, large-solid-angle, energy-dispersed, 
quasi-continuous along two spatial dimensions and highly efficient detection 
of high-energy electrons. This can be achieved, for the first time, 
by new solid-state pnCCD cameras, originally developed for X-ray detection 
at FELs and implemented into the CFEL-ASG-Multi-Purpose (CAMP) chamber 
developed by the Max-Planck Advanced Study Group (ASG) at the Center 
for Free Electron Laser science (CFEL) in Hamburg. The chamber along 
with the detectors have been described in detail before \cite{14} such that 
we will concentrate on the salient features and specific requirements 
to be fulfilled for electron detection. 

Due to the pixel size of $75$ $\mu$m $\times$ $75$ $\mu$m over a total area 
of $8$ cm $\times$ $8$ cm we can image the electrons in a quasi-continuous 
way along two spatial dimensions covering a large solid angle as requested. 
The high granularity of the detector simultaneously ensures, that large numbers 
of electrons, certainly $10^4$ per shot, can be recoded in an energy 
dispersed way which is feasible as long as each of the $10^6$ pixels 
is hit by not more than one electron. Moreover, the unprecedentedly high 
frame read-out rate of up to $200$ Hz is adapted to the maximum FEL 
repetition rate at LCLS of $120$ Hz, such that electrons can be recorded shot-by-shot. 

More care has to be taken concerning the energy resolution of the detector 
which is excellent, close to the theoretical limit (one-electron noise) 
for X-ray detection. Electrons, however, have to penetrate through a 
protecting ($50$ nm Al) as well a charge carrier, isolating depletion 
layer ($40$ nm SiO$_2$) meaning that only electrons at around $10$ keV 
can be detected with a probability of close to one and reasonable 
energy resolution, limited by the energy loss of the electrons in 
the entrance layers. Thus, electrons of $500$ eV to $2$ keV as considered 
to be ideal for the present purpose cannot be directly monitored but, rather, 
have to be post-accelerated towards the detector. 

\begin{figure}[t] 
\vspace{-0.5cm}
\begin{center}
\includegraphics[width=0.33\textwidth]{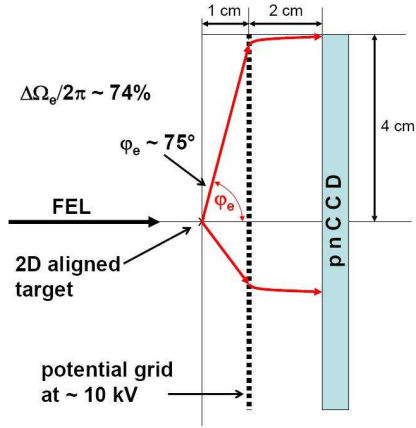}
\end{center} 
\vspace{-0.5cm} 
\caption{ Geometry for high-energy 
electron detection with the pnCCD in CAMP.}  
\label{figure5}  
\end{figure} 

\begin{figure}[t] 
\vspace{-4.5cm}
\begin{center}
\includegraphics[width=0.77\textwidth]{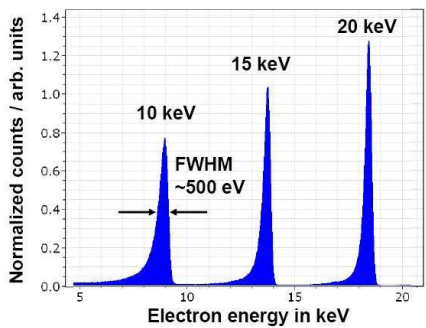}
\end{center} 
\vspace{-7.5cm} 
\caption{ Response of the pnCCD on electrons of different energies 
as indicated penetrating 40 nm SiO$_2$ insulating and 50 nm Al 
laser-light protection layers. }  
\label{figure6}  
\end{figure}

For that purpose, a potential grid on a voltage of $10$ kV 
will be placed about $1$ cm away from the interaction zone 
as depicted in figure \ref{figure5}, allowing to accept 
a solid angle $\Delta\Omega / 2 \pi \approx 0.74$ with 
the pnCCD ($8$ cm $\times$ $8$ cm) as close as $2$ cm 
downstream of the grid. Thus, the potential grid, 
serving for enhancing the solid angle, at 
the same time post-accelerates the electrons to $10.5$ keV or $12$ keV, respectively, 
enabling them to penetrate into the pnCCD through various protecting 
and carrier depletion layers. They are then detected with a FWHM of 
about $500$ eV as demonstrated in a recent test measurement shown 
in figure \ref{figure6} for electrons of different energies emerging from 
a electron microscope directly hitting (zero degree impact) the pnCCD. 
Thus, electrons with energies of around $1.7$ or $3.5$ keV emitted simultaneously 
in the present situation will end up with final energies of $11.7$ keV or $13.5$ keV, 
respectively, and, thus, can be easily discriminated against each other. 
In practice, that means that one can, in properly chosen situations, 
obtain holographic images simultaneously taken at two (several) photoelectron energies, 
illuminating the molecule from different sites from within. On the other hand, 
the electron energy resolution of $500$ eV clearly limits the method to cases 
where the photoelectrons emerging from different atoms in the molecule can 
be cleanly separated. Ideal cases are emitters with atomic numbers between 
chlorine ($Z = 17$) to gallium ($Z = 31$) as constituents of organic molecules 
containing essentially light atoms with $Z < 10$ with their K-shell energies 
(Cl: $2.8$ keV to Ga: $10.4$ keV) clearly separated from those of the lighter 
constituent atoms (C: $0.28$ keV, H: $0.014$ keV, up to F: $0.7$ keV) 
as well as from photoelectrons emerging from the $L$- and $M$-shells 
of the heavy species. Thus, even though the situation is not completely 
ideal, the large selection of constituent atoms for launching electrons 
from within allows one to certainly choose benchmark situations for 
chemical dynamics. Prominent examples are all (poly) pentene rings 
containing one Cl atom, easily attachable at different sites as emitter. 
Other molecules of interest for performing benchmarking, 
photoionization or dissociation experiments might 
be clorethylen \cite{45} or diphenylmethyl chloride \cite{46}. 

\section*{Summary and Outlook} 

We have developed a novel idea and provided model calculations 
for 'imaging molecules from within', which exploit the holographic diffraction 
pattern occurring when photoelectrons, site-specifically launched at certain atoms 
within molecules, are scattered off the three dimensional molecular potential 
on their way out. The observed patterns encode the $3D$ molecular pattern which 
is retrieved with {\AA}ngstr\"om resolution using reconstruction algorithms that 
have been developed for 'inside photoelectron holography' in solid state physics. 
Since femtosecond pulsed photon beams from X-ray free electron lasers are used 
to create the photoelectrons, the structure determination proceeds on an ultra-fast 
time scale. Along with the large number of photoelectrons that are produced due 
the high intensity of the FEL radiation, they are perfectly suited for probing 
the time-dependent structure of the object under investigation in pump-probe 
arrangements where the dynamics is induced either by an optical pump laser or 
by a replica of the FEL pulse itself as recently demonstrated at the FLASH \cite{47}. 

Different from any other scenario for coherent structure determination, 
we measure the three dimensional momentum vectors of high-energy electrons 
($E_e > 500$ eV) with large solid angle and energy dispersed exploiting 
recently developed large-area pnCCD detectors along with an electric field 
projection technique. Along with the unique short pulse properties of the FEL, 
this represents the ultimate technology to obtain real-time $3D$ structural 
information for small to moderate-size molecules by providing several unique features:

(i) As the outgoing electrons are diffracted in the combined electronic and 
nuclear charge distribution of the molecule, the diffraction pattern does 
not 'only' reveal information on the position of the nuclei but, moreover, 
on the complete multi-center molecular potential, i.e., the electronic orbitals. 
This is especially obvious when low-energy electrons are used to create 
the diffraction pattern. Thus, investigating the same reaction using photoelectrons 
of different energies, readily adjustable via tuning the FEL photon energies, 
it is expected (by comparison of the holographic images) to directly access 
the subtle interplay between electronic orbital and nuclear position dynamics 
beyond the Born-Oppenheimer approximation. 

(ii) By changing the photon energy and, thus, the energy and the de Broglie 
wavelength of the emerging electron, the latter can be 'adjusted' to 
any length-scale of interest for the respective molecule, possibly even to 
large-scale structures such as the folding patterns of proteins, 
in an 'inside-source holography' arrangement. Especially for larger proteins, 
it might, at some point, not be important to determine the position of each 
individual atom but, rather, to monitor the motion of groups of atoms with 
a 'fixed' structure relative to other groups.

(iii) Using high-energy photoelectron holography, it was demonstrated 
in solid state physics, that the positions of up to around $50$ nearest neighbor  
atoms could be retrieved. Even though this seems not to be really much if 
the dynamics in a large molecule e.g. a protein shall be explored, it might 
be an ideal method to trace the decisive early-time dynamics around a photo receptor site.

(iv) While pump-probe experiments that study the time-dependent evolution 
of molecular valence orbitals have recently been performed with high-harmonic 
laser sources \cite{41}, our approach using shorter wavelength probe pulses allows 
accessing localized inner-shell electrons, which is essential for structural 
determination via electron diffraction since it allows launching 
the photoelectron from specific sites within the molecule. 

(v) Due to the possibility to simultaneously measure both ions 
and high-energy electrons in CAMP, we will be able to continuously 
monitor the degree of alignment as well as the fragmentation channels 
(via measuring the fragment kinetic energies) in the pump-probe experiments 
while taking the electron diffraction data. This (quasi-) coincidence 
mode may also allow cleaning up the experimental data in the post-analysis,  
especially if covariant mapping methods \cite{48} are applied. 

(vi) Since the core-holes created as a consequence of photoelectron emission 
decay either via emission of Auger electrons or of fluorescence photons, 
it should be possible (if the various energies can be separated in the pnCCD) 
to also record holographic images for these reaction products, as demonstrated 
as well in solid state physics. Auger decays typically take place on a time-scale of about 
$10$ fs, i.e. within the present FEL pulse length, and may, in certain cases, 
directly influence the photoelectron emission. Radiative transitions in medium 
heavy atoms on the other hand are one to two orders of magnitude slower 
and, accordingly, would not affect the angular distribution 
of the directly emitted photo electron. 

(vii) Since the intensities required for the X-ray probe pulse are rather 
small (large focal diameters) compared to X-ray diffraction studies 
(due to the fact that photoionization 
cross sections close to an edge are several orders of magnitude larger than 
X-ray scattering cross sections), radiation damage during the pulse is negligible. 
Furthermore, this makes photoelectron diffraction a viable tool to study ultra-thin 
ensembles, such as beams of conformer-selected molecules,  
Stark decelerated (trapped) molecules with typical target densities of 
only about $10^6$ cm$^{-3}$ or ultra-cold (molecular) ensembles  
in Bose-Einstein condensates or degenerate Fermi gases. To the best 
of our knowledge, structural dynamics of this rather large class of 
samples will not be accessible by any other present technique.

\section*{Acknowledgement} 

Support from the Max-Planck Advanced Study Group at CEFL is gratefully
acknowledged. BN would like to thank DFG for support under the project
VO 1278/2-1. We would like to thank Sascha Epp, Artem 
Rudenko, Simone Techert, Robert Moshammer, Ilme Schlichting, 
Robert Hartmann, Regine de Vivie-Riedle and Henrik Stapelfeldt for helpful discussions.

\end{document}